\documentclass[twocolumn]{aastex631}
\usepackage{graphicx}
\usepackage{hyperref}
\usepackage{mathtools}
\usepackage{xspace}

\usepackage{color}

\def\eg{{\it e.g., }}

\def\ie{{\it i.e., }}

\def\gsim{~\rlap{$>$}{\lower 1.0ex\hbox{$\sim$}}}
\def\lsim{~\rlap{$<$}{\lower 1.0ex\hbox{$\sim$}}}

\def\tess{{\it TESS}\xspace}

\def\msun{{M_\odot}}

\def\mearth{{M_\oplus}}
\def\rearth{{R_\oplus}}

\def\vplanet{\texttt{\footnotesize{VPLanet}}\xspace}
\def\atmesc{\texttt{\footnotesize{AtmEsc}}\xspace}

\def\distorb{\texttt{\footnotesize{DistOrb}}\xspace}

\def\eqtide{\texttt{\footnotesize{EqTide}}\xspace}

\def\magmoc{\texttt{\footnotesize{MagmOc}}\xspace}

\def\radheat{\texttt{\footnotesize{RadHeat}}\xspace}
\def\spinbody{\texttt{\footnotesize{SpiNBody}}\xspace}
\def\stellar{\texttt{\footnotesize{STELLAR}}\xspace}
\def\thermint{\texttt{\footnotesize{ThermInt}}\xspace}

\def\nbody6{\texttt{\footnotesize{NBODY6++}}\xspace}

\def\k40{{$^{40}$K}}
\def\th232{{$^{232}$Th}}
\def\ur235{{$^{235}$U}}
\def\ur238{{$^{238}$U}}
\def\co2{{CO$_2$}}

\def\apjs{{\it Astrophys.~J.~Supp.}}

\def\apjl{{\it Astrophys.~J.~Lett.}}

\def\aap{{\it Astron.~\&~Astrophys.}}

\begin{document}
\title{History and Habitability of the LP 890-9 Planetary System}


\correspondingauthor{Rory Barnes}
    \email{rory@astro.washington.edu}
\author[0000-0001-6487-5445]{Rory Barnes}
    \affiliation{Astronomy Department, University of Washington, Box 315180, Seattle, WA 98195}

\author[0000-0002-8341-0376]{Laura N. R. do Amaral}
     \affiliation{Instituto de Ciencias Nucleares,
               Universidad Nacional Autónoma de México, Cto. Exterior S/N, C.U., Coyoacán, 04510 Ciudad de México, CDMX}

\author[0000-0002-7961-6881]{Jessica Birky}
     \affiliation{Astronomy Department, University of Washington, Box 315180, Seattle, WA 98195}

 \author{Ludmila Carone}
     \affiliation{Space Research Institute, Austrian Academy of Sciences, Schmiedlstrasse 6, A-8042 Graz, Austria}
     \affiliation{Centre for Exoplanet Science, School of Physics \& Astronomy, University of St Andrews, North Haugh, St Andrews, KY169SS}

 \author{Peter Driscoll}
     \affiliation{Earth and Planets Laboratory, Carnegie Institution for Science, 5241 Broad Branch Rd, Washington, DC, 20015, USA}

\author[0000-0003-3888-3753]{Joseph R. Livesey}
     \affiliation{Astronomy Department, University of Washington, Box 315180, Seattle, WA 98195}
     \affiliation{Department of Astronomy, University of Wisconsin--Madison, 475 N. Charter Street, Madison, WI 53706, US}

\author{David Graham}
     \affiliation{Institut für Theoretische Astrophysik (ITA), Zentrum für Astronomie (ZAH), Universität Heidelberg, Albert-Ueberle-Str.2 69120 Heidelberg, Germany}

\author[0000-0002-7733-4522]{Juliette Becker}
\affiliation{Division of Geological and Planetary Sciences, California Institute of Technology, Pasadena, CA 91125, USA}
\affiliation{Department of Astronomy, University of Wisconsin--Madison, 475 N. Charter Street, Madison, WI 53706, US}

\author[0000-0003-1535-5587]{Kaiming Cui}
\affiliation{Tsung-Dao Lee Institute, Shanghai Jiao Tong University, 800 Dongchuan Road, Shanghai 200240, People’s Republic of China}

\author[0000-0001-8355-2107]{Martin Schlecker}
\affiliation{Steward Observatory and Department of Astronomy, The University of Arizona, Tucson, AZ 85721, USA}
    
\author{Rodolfo Garcia}
    \affiliation{Astronomy Department, University of Washington, Box 315180, Seattle, WA 98195}

\author{Megan Gialluca}
    \affiliation{Astronomy Department, University of Washington, Box 315180, Seattle, WA 98195}

\author[0000-0002-7139-3695]{Arthur Adams}
\affiliation{Department of Earth and Planetary Sciences, University of California, Riverside, CA, 92521, USA}

\author[0000-0001-7483-9982]{MD Redyan Ahmed}
     \affiliation{Department of Physical Sciences, Indian Institute of Science Education and Research Kolkata, West Bengal,741246, India }

\author{Paul Bonney}
    \affiliation{Department of Physics, University of Arkansas, 825 W. Dickson St., Fayetteville, Arkansas 72701, USA}

\author{Wynter Broussard}
\affiliation{Department of Earth and Planetary Sciences, University of California, Riverside, CA, 92521, USA}

\author[0000-0001-5408-4156]{Chetan Chawla} 
    \affiliation{Independent Researcher, New Delhi - 110027, India}

\author[0000-0001-9984-4278]{Mario Damasso}
\affiliation{INAF -- Osservatorio Astrofisico di Torino, Via Osservatorio 20, I-10025, Pino Torinese (TO), Italy} 

\author[0000-00002-9209-58320]{William C. Danchi}
\affiliation{NASA Goddard Space Flight Center, Exoplanets and Stellar Astrophysics Laboratory, Greenbelt, MD 20771}

\author{Russell Deitrick}
\affiliation{School of Earth and Ocean Sciences, University of Victoria, Victoria, British Columbia, Canada}

\author[0000-0002-7008-6888]{Elsa Ducrot}
\affiliation{AIM, CEA, CNRS, Universit\'e Paris-Saclay, Universit\'e de Paris, F-91191 Gif-sur-Yvette, France }
\affiliation{Paris Region Fellow, Marie Sklodowska-Curie Action}

\author[0000-0002-3099-0493]{Emeline F. Fromont}
\affiliation{Department of Astronomy, University of Maryland, College Park, MD 20742, USA}

\author[0000-0003-4224-6829]{Brandt A. L. Gaches}
    \affiliation{Department of Space, Earth \& Environment, Chalmers University of Technology, SE-412 96 Gothenburg, Sweden}
    \affiliation{Center for Planetary Systems Habitability, The University of Texas at Austin, Austin, TX 78712, USA.}

\author[0000-0002-8920-1495]{Sakshi Gupta}
\affiliation{Center of Excellence in Space Sciences India, Indian Institute of Science Education and Research Kolkata, Mohanpur 741246, India}
\affiliation{Department of Physical Sciences, Indian Institute of Science Education and Research Kolkata, Mohanpur 741246, India}

\author[0000-0002-0139-4756]{Michelle L. Hill}
\affiliation{Department of Earth and Planetary Sciences, University of California, Riverside, CA, 92521, USA}

\author{James A. G. Jackman}
\affiliation{School of Earth and Space Exploration, Arizona State University, Tempe, AZ, 82587, USA}

\author[0000-0003-0475-8479]{Estelle M. Janin}
    \affiliation{School of Earth and Space Exploration, Arizona State University, Tempe, AZ, 82587, USA}

\author[0000-0002-1505-287X]{Mikołaj Karawacki}
\affiliation{Department of Physics, Astronomy and Applied Informatics, Nicolaus Copernicus University, Toruń, Poland}

\author[0000-0002-9705-8019]{Matheus Daniel Koren}
    \affiliation{Instituto de Matemática, Estatística e Física, Universidade Federal do Rio Grande, Rio Grande, RS, Brazil}

\author{Roberto La Greca}
    \affiliation{Biology Consultant Freelancer, Lodi CAP 26900, Lombardia, Italy}

\author[0000-0003-1906-5093]{Michaela Leung} 
\affiliation{Department of Earth and Planetary Sciences, University of California, Riverside, CA, 92521, USA}

\author[0000-0002-2857-0131]{Arturo Miranda-Rosete}
    \affiliation{Instituto de Ciencias Nucleares,
              Universidad Nacional Autónoma de México, Cto. Exterior S/N, C.U., Coyoacán, 04510 Ciudad de México, CDMX}

\author[0000-0001-5489-540X]{Michael Kent A. Olohoy}
\affiliation{Department of Earth and Space Sciences, Rizal Technological University, Boni Ave, Mandaluyong, 1550 Metro Manila, Philippines}

\author{Cecelia Ngo}
\affiliation{Department of Physics and Astronomy, University of California, Riverside, 92521, USA}

\author[0000-0001-6289-0756]{Daria Paul}
    \affiliation{Physikalisches Institut, Universit{\"a}t zu K{\"o}ln, Z{\"a}lpicher Stra{\ss}e 77, 50937, K{\"o}ln, Germany}

\author[0000-0003-1057-2320]{Chandan Kumar Sahu}
    \affiliation{National Institute of Science Education and Research Bhubaneswar, Jatni 752050, Odisha, India}

\author [0000-0002-2827-579X] {Debajyoti Basu Sarkar}
\affiliation{Independent Researcher, White Hall, AR 71602, USA} 

\author[0000-0002-0797-5017]{Mohammad Afzal Shadab}
    \affiliation{Department of Civil and Environmental Engineering, Princeton University, Princeton, NJ 08544, USA.}     
    \affiliation{Oden Institute for Computational Engineering and Sciences, The University of Texas at Austin, Austin, TX 78712, USA.}    
    \affiliation{Center for Planetary Systems Habitability, The University of Texas at Austin, Austin, TX 78712, USA.}

\author[0000-0002-2949-2163]{Edward W. Schwieterman}
\affiliation{Department of Earth and Planetary Sciences, University of California, Riverside, CA, 92521, USA}

\author{Melissa Sedler}
    \affiliation{Independent Researcher, Phoenix, AZ, 85048, USA}

\author{Katie Texeira}
\affiliation{Department of Astronomy, University of Texas at Austin, Austin, TX 78712, USA}

\author[0000-0001-9504-3174]{Allona Vazan}
    \affiliation{Astrophysics Research Center (ARCO), Department of Natural Sciences, The Open University of Israel, Raanana 4353701, Israel}

\author{Karen N. Delgado Vega}, 
    \affiliation{University of Puerto Rico at Arecibo}

\author[0000-0002-4258-7175]{Rohit Vijayakumar}
    \affiliation{Divecha Center for Climate Change, India Institute of Science, Bengaluru 560012, Karnataka, India}
    
\author[0000-0003-1896-7597]{Jonathan T. Wojack} 
\affiliation{University of Southern Queensland, Toowoomba QLD 4350, Australia}

\keywords{astrobiology - methods: numerical - software: simulations - planets and satellites: individual: LP 890-9: dynamical evolution and stability - planets-star interactions}

\begin{abstract}
     We present numerous aspects of the evolution of the LP 890-9 (SPECULOOS-2/TOI-4306) planetary system, focusing on the likelihood that planet c can support life. We find that the host star reaches the main sequence in 1 Gyr and that planet c lies close to the inner boundary of the habitable zone. We find the magma ocean stage can last up to 50 Myr, remove 8 Earth-oceans of water, and leave up to 2000 bars of oxygen in the atmosphere. However, if the planet forms with a hydrogen envelope as small as 0.1 Earth-masses, no water will be lost during the star's pre-main sequence phase from thermal escape processes. We find that the planets are unlikely to be in a 3:1 mean motion resonance and that both planets tidally circularize within 0.5 Gyr when tidal dissipation is held constant. However, if tidal dissipation is a function of mantle temperature and rheology, then we find that planet c's orbit may require more than 7 Gyr to circularize, during which time tidal heating may reach hundreds of terawatts. We thus conclude that the habitability of planet c depends most strongly on the initial volatile content and internal properties, but no data yet preclude the viability of an active biosphere on the planet. 
\end{abstract}

\section{Introduction} 

The  LP 890-9 system consists of a $0.12\msun$ star and two approximately Earth-radius planets \citep{Delrez22}, with the outer planet in or near the circumstellar habitable zone \citep[HZ; ][]{Dole64,Kasting93,Kopparapu13}. The planetary system was discovered via transits from both \tess and ground-based telescopes, but radial velocity data failed to achieve a detection so the masses are very poorly constrained. At a distance of 32 pc, planet c is in the fourth closest system known to harbor at least one potentially habitable planet that transits its host star, after TRAPPIST-1 \citep{Gillon17}, LHS 1140 \citep{Dittmann17} and Gl 12 \citep{Dholakia24,Kuzuhara24}. At this distance, both planets' atmospheres may be amenable to follow-up spectroscopy with the James Webb Space Telescope (JWST)~\citep{Delrez22}. With this possibility in mind, we have performed a set of simulations that provide preliminary insight into the history and habitability of the LP 890-9 system. 

We cast a wide net in our investigations, considering phenomena ranging from the geodynamo in the planetary cores to the long term evolution of the orbits. The observed and derived properties relevant to this study, all taken from \cite{Delrez22}, are presented in Tables 1 and 2. For the non-linear and high-dimensional problem of planetary system evolution, these sets of parameters are insufficient to accurately constrain any realistic model of planet c's habitability \citep{MeadowsBarnes18}. We therefore augment these sets of priors with numerous other assumptions and simplifications, described and justified in $\S$2, to predict aspects of the orbital, rotational, atmospheric, surface, and internal evolution of the star and planets.

\begin{table}
    \centering
    Table 1: Observed and Adopted Properties of LP 890-9\\
    \begin{tabular}{lcl}
        \hline
        Parameter & Value & Description\\
        \hline
        $L_{bol}$ ($10^{-3} L_\odot$) & $1.44\pm0.04$ & Total luminosity\\
        $T_{eff}$ (K) & $2850\pm75$ & Effective temperature\\
        $P_{rot}$ (d) & $<100$ & Rotation period\\
        log($L_{H\alpha}/L_{bol}$) & $-4.6\pm0.1$ & Normalized H$\alpha$ luminosity\\
        $[$Fe/H$]$ (dex) & $-0.03\pm0.09$ & Metallicity\\
        $M_*$ ($M_\odot$) & 0.118$\pm$0.002 & Stellar Mass\\
        $R_*$ ($R_\odot$) & 0.1556$\pm$0.0086 & Stellar Radius\\
        Age (Gyr) & $7.2^{+2.2}_{-3.1}$ & Age\\
        $L_{XUV}/L_{bol}$ & $(5\pm2)\times10^{-5}$ & Adopted XUV luminosity\\
        \hline\hline
    \end{tabular}
\end{table}

\begin{table*}
    \centering
    Table 2: Observed and Adopted Properties of the LP 890-9 Planets
    \begin{tabular}{lccl}
        \hline
        Parameter & b & c & Description\\
        \hline
        $m_p$ ($\mearth$) & $<13.2$ & $<25.3$ & RV upper limits\\
        $m_p$ ($\mearth$) & $2.3^{+1.7}_{-0.7}$ & $2.5^{+1.8}_{-0.8}$ & Adopted mass \\
        $r_p$ ($\rearth$) & $1.320^{+0.053}_{-0.027}$ & 1.367$^{+0.055}_{-0.039}$ & Planetary radius\\
        $P_{orb}$ (d) & 2.7299025$\pm3\times10^{-6}$ & 8.4575$\pm3\times10^{-4}$ & Orbital period\\
        $I (^\circ$) & 89.67$^{+0.22}_{-0.33}$ & 89.287$^{+0.026}_{-0.047}$ & Inclination w.r.t. sky plane\\
        \hline\hline
\end{tabular}
\end{table*}

In the next section we describe the physical models we used to simulate the system, as well as the initial conditions for our simulations. In $\S$3, we present the results for stellar evolution ($\S$3.1), atmospheric escape ($\S$3.2), orbital evolution ($\S$3.3), magma ocean phase ($\S$3.4), and finally the post-magma ocean thermal and magnetic evolution ($\S$3.5). We then interpret the results in $\S$4 and conclude in $\S$5. Code to generate the figures is publicly available\footnote{https://github.com/RoryBarnes/LP890-9}.

\section{Methods}

To simulate this system, we used \vplanet\footnote{Publicly available at https://github.com/VirtualPlanetaryLaboratory/VPLanet.}, a multi-purpose planetary evolution code  that breaks down the physical and chemical processes of planetary evolution into ``modules'' that can often be coupled together to track feedbacks \citep{Barnes20}. In this section we first briefly review the physics employed in this study, and refer the reader to \cite{Barnes20} and \vplanet's online documentation for detailed discussions of the modules and their implementation. We then describe the details of the individual experiments we conducted.

\subsection{Physical Model}\label{subsec:PhysicalModel}

The host star primarily evolves according to a bicubic interpolation of the \cite{Baraffe15} model grids. These models predict (as a function of mass, age, and metallicity), the radius, bolometric (total) luminosity $L_{bol}$, effective temperature, and mass concentration. The XUV luminosity follows the \cite{Ribas05} model, in which the initial, or ``saturated,'' XUV luminosity ($f_{sat}$) remains constant for 0.1--8 Gyr, a duration called the saturation time ($t_{sat}$). The XUV luminosity, $L_{XUV}$, then decays by a power law described by a parameter we call $\beta_{XUV}$ \citep[see also][]{Peacock20,Johnstone21,RicheyYowell22}. In other words,
\begin{align}
\label{eq:lxuv}
\frac{L_\mathrm{XUV}}{L_\mathrm{bol}} = \left\{
				\begin{array}{lcr}
					f_\mathrm{sat} &\ & t \leq t_\mathrm{sat} \\
					f_\mathrm{sat}\left(\frac{t}{t_\mathrm{sat}}\right)^{-\beta_\mathrm{XUV}} &\ & t > t_\mathrm{sat}.
				\end{array}
				\right.
\end{align}
Note that the position of the HZ is only a function of $L_*$ and $T_{eff}$ \citep{Kopparapu13}. LP 890-9 is a low mass star, and therefore early in its lifetime its luminosity decreased slowly over hundreds of Myr while its temperature remained approximately constant \citep{Hayashi61}. 

We explore different starting times (5--20 Myr) of each simulation to account for variations in the timescale for planet formation \citep[see, for example,][]{Raymond07,Lambrechts19,Clement22,RaymondMorbidelli22}. In practice, this uncertainty is manifested in the starting time in the Baraffe et al. grids. We don't begin any simulation earlier than 5 Myr into these grids in order to avoid any complications due to the initial physics of infall as well as instabilities in the bicubic interpolation method, which relies on numerous data points in both the time and mass dimensions. Thus, the ``times'' reported below are 5--20 Myr less than the values reported in the Baraffe et al. grids. More information about the \stellar module can be found in \cite{Barnes20}, Appendix J.

The planetary orbits are influenced by both tidal interaction with the host star and the exchange of angular momentum between the planets due to gravitational interactions. For the long-term orbital evolution of the system, we use a constant-time-lag version of the equilibrium tide model \citep{Darwin1880,GoldreichSoter66,FerrazMello08,Leconte10}, which reduces the tidal effects to two parameters: the tidal time lag $\tau$ (which is analogous to the tidal $Q$) and the Love number $k_2$. In general, tidal effects will tend to decrease the eccentricity $e$ and semi-major axis $a$ \citep{Barnes17}. The details of the \eqtide module can be found in \cite{Barnes20}, App.~E.

For the planet-planet interactions, we employ two different models depending on the goal. To examine the possibility that the planets are in a mean motion resonance (MMR), we use an N-body model, the \spinbody module of \vplanet, that calculates the forces directly between the star and two planets. A complete description of the \spinbody module can be found in \cite{Barnes20}, App.~I.

The ratio of the orbital periods between planets c and b is 3.1, suggesting the planets are near, but not in, a 3:1 MMR. Recent work has shown that for first order MMRs, \eg 2:1, a separatrix exists at period ratios about 0.6\% wider than perfect resonance \citep{GoldbergBatygin22}, but no previous work has yet explored second order resonances. As such an analysis is beyond the scope of the current study, we instead examine the likelihood of resonance the classical way: by searching for librations of the resonant arguments. For this system, three eccentricity-type resonance arguments, $\theta$ exist:
\begin{equation}\label{eq:resarg1}
\theta_1 = 3\lambda_c - \lambda_b - 2\varpi_b,
\end{equation}
\begin{equation}\label{eq:resarg2}
\theta_2 = 3\lambda_c - \lambda_b - 2\varpi_c,
\end{equation}
and
\begin{equation}\label{eq:resarg3}
\theta_3 = 3\lambda_c - \lambda_b - \varpi_c - \varpi_b,
\end{equation}
where $\lambda$ is the mean longitude and $\varpi$ is the longitude of periastron. 

For long-term (Gyr) integrations, we use \texttt{\vplanet's} 4th order ``secular model'' (\distorb) that assumes the long-term orbital evolution is well-modeled by torques between elliptical rings of matter \citep{Ellis2000,Deitrick18a}. This method is much more computationally efficient than direct calculations, but its accuracy fades as the eccentricity and relative inclination become larger than 0.3 and 25$^\circ$, respectively. The interplanetary gravitational forces will tend to drive oscillations in $e$ and $i$, where $i$ is the inclination relative to a reference plane. In this study we set the reference plane to be the invariable (or fundamental) plane, \ie the plane perpendicular to the total angular momentum vector. The \distorb module is described in detail in \cite{Barnes20}, App.~C.

The long-term orbital evolution of the system depends on one conservative and one dissipative model. This interplay results in many parameters evolving like damped-driven harmonic oscillators \citep{MurrayDermott99}. One (or more) frequency's amplitude effectively damps to zero and the system reaches an equilibrium state. For planetary systems, the equilibrium state is typically  planets with fixed non-zero eccentricities and major axes that precess in unison \citep{WuGoldreich02}.

Moving on to the planets themselves, we will consider two types of atmospheres: hydrogen-rich worlds, which we will call ``mini-Neptunes'' and hydrogen-free worlds, which we will call ``super-Earths'' \citep{Barnes09}. For the mini-Neptune case, we assume the planet can lose hydrogen via Roche lobe overflow \citep{OwenWu17}, radiation/recombination-limited escape \citep{MurrayClay09}, or energy-limited escape \citep{Watson81,Erkaev07,Luger15}. We calculate the method of loss based on instantaneous conditions, \ie as the XUV radiation evolves, the manner in which the hydrogen is lost is automatically adjusted during the integration \citep{Amaral22}. We employ \vplanet's \atmesc module to perform these calculations, which is described in \cite{Barnes20}, App.~A. Note that this model does not include any effects of molecular hydrogen on the climates of these planets \citep{PierrehumbertGaidos11}.

We assume super-Earths possess large initial abundances of water that erode away due to photolysis followed by H escape \citep{Watson81,Chassefiere96,LugerBarnes15}. Our model assumes loss is initially energy-limited, but if a sufficient amount of liberated oxygen accumulates in the atmosphere, the escape becomes diffusion-limited. We assume the efficiency of the transformation of photon energy to the kinetic energy of escaping protons scales with the incident XUV flux \citep{Bolmont17}, and is generally around 10\%. 
We further assume that water is only photolyzed after all of a primordial hydrogen envelope has escaped in our model and when the planet is interior to the HZ.

Water photolysis also releases free oxygen atoms. We consider two possibilities for the fate of this liberated oxygen: 1) it accumulates in the atmosphere, where it may be dragged to space by escaping hydrogen atoms when the XUV flux is large enough \citep{Hunten87,LugerBarnes15}, or 2) it all reacts with the surface and enters the mantle. Thus, in some cases, we expect oxygen to permanently accumulate in the atmosphere.

For this study, we assume the planetary compositions are Earth-like, which may or may not be the case. We make this choice because of the relatively poor constraints on the material properties of other abundance patterns, such as larger abundances of magnesium or calcium \citep[see \eg][]{Bond10}. Moreover, \vplanet's post-magma ocean interior model is only calibrated to Earth and Venus, so we limit our scope to their composition.

We consider the possibility that the planets formed very hot with a fully molten mantle, as is likely for Earth right after the Moon-forming impact \citep{Stevenson87}. During this ``magma ocean'' phase, the volatile gases are partitioned between the interior and atmosphere in a pseudo-equilibrium \citep{Abe97,elkinstanton2008b,Schaefer16} (the escaping atmosphere and secular cooling of the planet prevent actual equilibrium). Oxygen can bond with iron in the mantle to form the solid Fe$_2$O$_3$, effectively removing O from the atmosphere. As the mantle solidifies from the core/mantle boundary to the surface, water can become trapped in the mantle. To simulate the magma ocean phase, we primarily rely on \vplanet's \magmoc module, which is described in \cite{Barth21}.

Once the mantle's melt fraction drops to a sufficiently low value, the viscosity decreases and the equilibrium between the interior and atmosphere breaks \citep{Barth21}. This transition marks the end of the magma ocean phase and the beginning of either a plate tectonics or stagnant lid thermal/volatile/magnetic evolution \citep{DriscollBercovici13,DriscollBercovici14}, which may still be affected by tidal heating \citep{Jackson08b,Jackson08c,Henning09,Barnes13,DriscollBarnes15}. We evaluate the post-magma ocean evolution of the planets with \vplanet's \thermint module, which is described in \cite{Barnes20}, App.~K. Specifically, we assume the planets evolve in a plate tectonics mode that cools the interior more rapidly than a stagnant lid planet. This choice should increase the temperature difference between the core and mantle, promoting core convection and a geodynamo. We also note that, following \cite{DriscollBercovici13}, we assume a modest amount of potassium heating in the core, which can resolve the so-called new core paradox and thermal catastrophes implied by thermal models that do not include deep heating or an insulating layer \citep{Korenaga06,LabrosseJaupart07}.

The above set of models provides preliminary insight into multiple aspects of the planetary system's evolution. Next we describe the parameter spaces we explored.

\subsection{Initial Conditions}

We performed 1000 simulations of the stellar evolution with each individual system's initial conditions sampled as a Gaussian according to the values in Table 1. Furthermore, we simulated each system to a current age that is sampled from the Gaussian distribution within uncertainties listed in Table 1. In practice, this latter choice does not affect the results. For the XUV evolution, we assumed the following initial distributions of parameters:  $-3.5 \le \log_{10}f_{sat} \le -2.5$; 1 Gyr $\le t_{sat} \le$ 10 Gyr; and $-2 \le \beta < 0$ based on previous results for low mass M dwarfs \citep{Peacock20,Magaudda20,Johnstone21,Birky21}. These simulations start at 5 Myr of stellar evolution, which planet formation models suggest is about the minimum for how long it would take for these planets to fully form \citep{Raymond07,Lambrechts19,Clement22}.

We assume that the planet is fully covered by water, and some cases possess a primordial envelope made of hydrogen.  We considered 3 different values of initial surface water of 1, 3, and 10 TO\footnote{A ``terrestrial ocean,'' the mass of water in modern Earth's oceans, equal to $1.39\times 10^{21}$ kg.}, and initial primordial hydrogen envelope masses of 0.001, 0.01, and 0.1 M$_{\oplus}$.

Moving on to the orbital evolution, we performed a suite of N-body simulations designed to evaluate the possibility that the planets are in the 3:1 resonance. Each case was integrated for 10-1000 years, or about 400-40,000 orbits of the outer planet. We found this duration was sufficient to identify librating resonant arguments. We varied both planets' mean anomalies between 0 and 360$^\circ$ in increments of 60$^\circ$; both planets' eccentricities between 0.05 and 0.3 in increments of 0.05, and the outer planet's longitude of pericenter between 0 and 360$^\circ$ in increments of $60^\circ$. Note that 0.3 is the maximum eccentricity that does not cause orbit crossing at the current semi-major axes.

To model the long-term evolution of the orbital elements, the initial eccentricities of each planet were sampled uniformly over $[0, 0.3]$. The initial longitudes of ascending node and periastron were sampled uniformly in the range $[0, 360^\circ)$. In this parameter space, we find that the planets circularize in $\sim$100 Myr, see $\S$3.3, so we don't simulate for longer timescales.

For the magma ocean phase, we explored solely the evolution of LP 890-9 c because \vplanet's \magmoc{}  model is currently only validated for planets in the HZ of their host star. We considered the 1$\sigma$ lower and upper limit of the planet mass, respectively, and investigated initial water masses between 1 and 100 TO with 20 simulations and logarithmic scaling.

To assess the importance of long-term radiogenic and tidal heating, we ran a suite of integrations with a range of initial physical parameters. We chose planet masses within the 1$\sigma$ estimate given in \citet{Delrez22}, initial eccentricities ranging between 0 and 0.5 for each planet and mantle viscosity activation energies between 2.75$\times 10^5$ and 3.5$\times 10^5$, corresponding to tidal $Q$s between 10 and 200. For comparison, Earth's mantle viscosity activation energy $A_\nu = 3\times 10^5$. The grid spanned eight values in each dimension, for a total of 512 integrations. In all cases, we assumed Earth's abundances of the radiogenic isotopes $^{40}$K, $^{232}$Th, $^{235}$U, and $^{238}$U.

\section{Results}

\subsection{Stellar Evolution}

\begin{figure*}
    \centering
    \includegraphics[width=\linewidth]{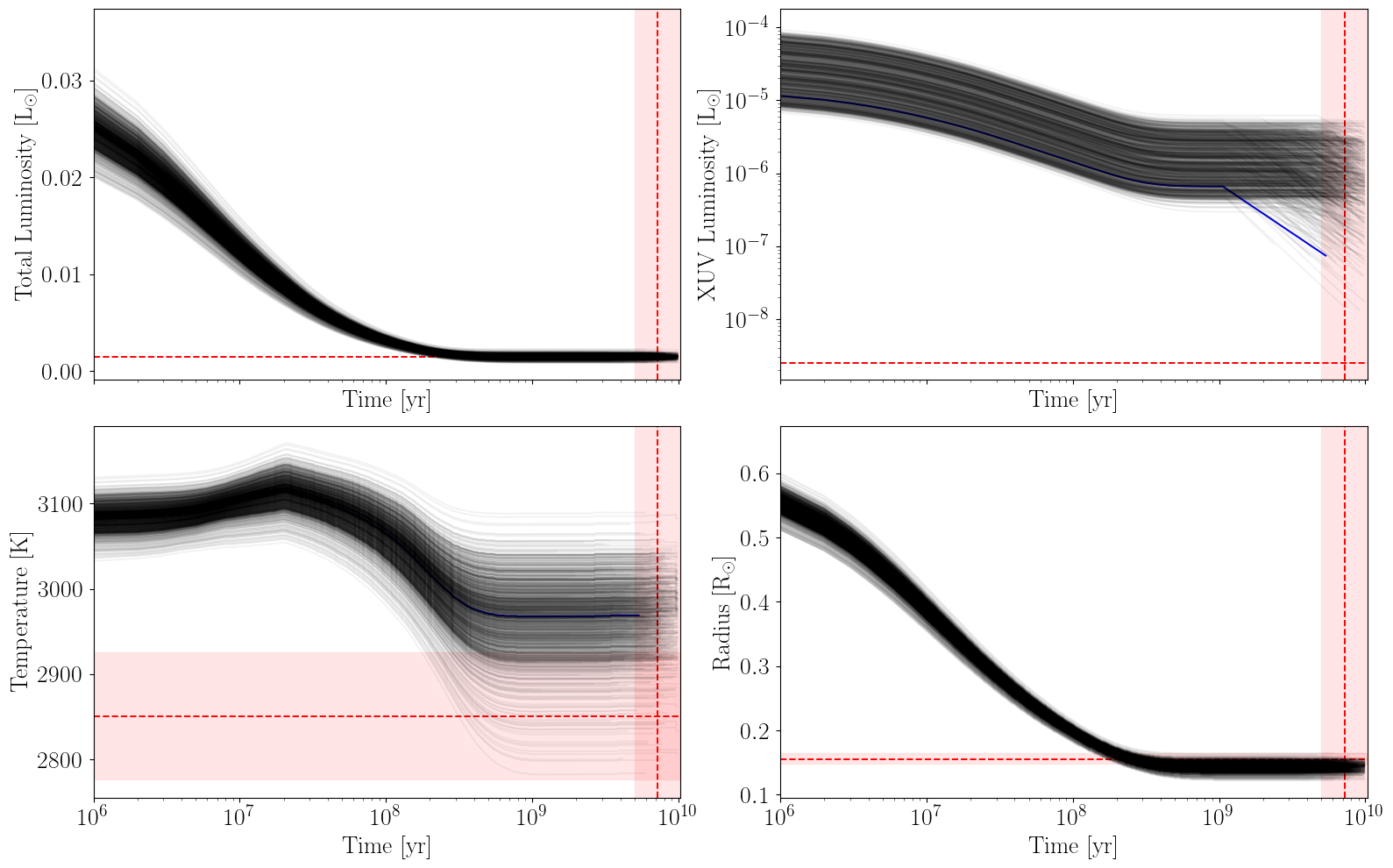}
    \caption{Evolution of some of LP 890-9's stellar properties. Note that the ``Time'' in each panel is offset by 5 Myr from the \cite{Baraffe15} model grids that form the foundation for the stellar evolution model, see $\S$2.1. Each panel includes 10\% of 1000 trials sampled within the ranges described in the text. The red lines and shading represent the best fit and 1-$\sigma$ uncertainties of each parameter, if available. {\it Top left:} Total bolometric luminosity. {\it Top right:} XUV luminosity. The blue curve is the best fit history of the star. {\it Bottom left:} Effective temperature. {\it Bottom right:} Radius.}
    \label{fig:StellarEvol}
\end{figure*}

In Fig.~\ref{fig:StellarEvol} we show the evolution of the star's bolometric luminosity, XUV luminosity, effective temperature, and radius. While the Baraffe et al. tracks reproduce the observed luminosity quite well, they tend to be displaced by about 2 standard deviations from the best fit values of $T_{eff}$ and radius. This result is consistent with previous analyses that found that M dwarf radii are systematically larger than theoretical predictions \citep[\eg][]{MorrellNaylor19}. However, for the purposes of our investigation, this offset is acceptable since the HZ is only weakly dependent on effective temperature, and tidal evolution of the planets will be dominated by dissipation in the planets \citep{MardlingLin02}.

\begin{figure}
    \centering
    \includegraphics[width=\linewidth]{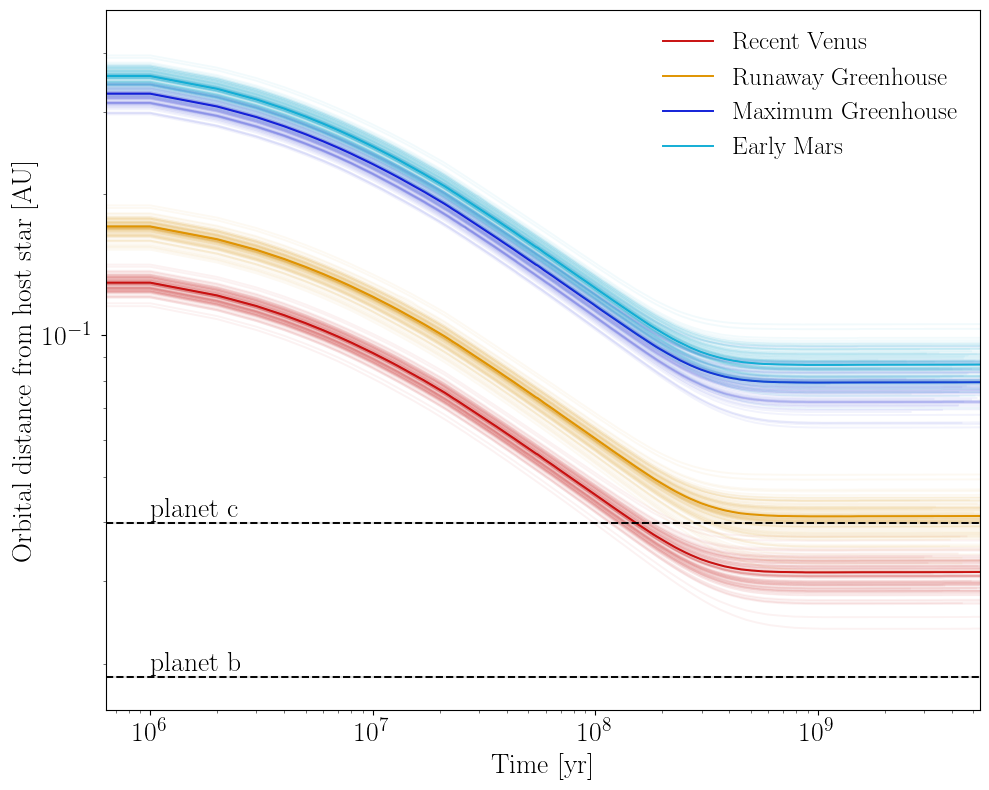}
    \caption{Evolution of LP 890-9's HZ for 10\% of the 1000 trials. The pale blue curves correspond to the ``early Mars'' limit, dark blue to ``maximum greenhouse,'' orange to ``runaway greenhouse'' and red to ``recent Venus'' \citep{Kopparapu13}. Thin curves represent individual trials, while the thick line is the average. The current orbits of planets b and c, assuming the best fit stellar mass, are shown with dashed black lines.}
    \label{fig:HZ}
\end{figure}

Figure \ref{fig:HZ} shows the evolution of the HZ for the range of plausible stellar masses. The limits are taken from \cite{Kopparapu13}. While planet b is always well interior to the HZ, planet c is within the ``optimistic'' HZ for all plausible stellar masses, and could be in the ``conservative'' HZ. Thus, planet c probably spent between 100 and 600 Myr interior to the HZ.

\subsection{Atmospheric Evolution}
\label{sec:AtmoEvol}

\begin{figure*}
    \centering
    \includegraphics[width=0.9\linewidth]{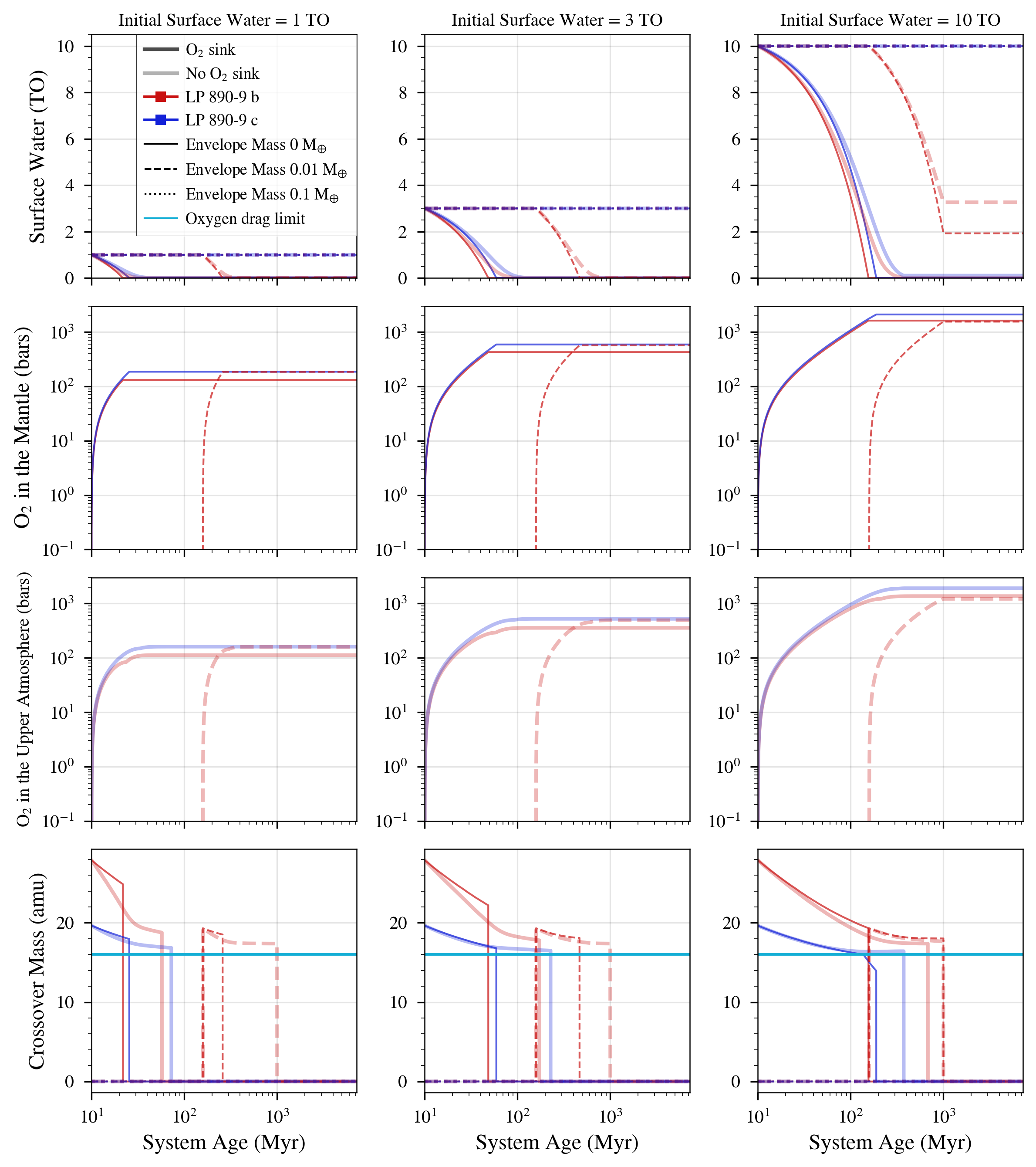}
    \caption{Evolution of the atmospheric parameters for LP 890-9 b (red) and c (blue), with an initial surface water content of 1 TO (left panels), 3 TO (middle panels), and 10 TO (right panels). {\it First row:} Surface water content. The solid, dashed, and dotted curves represent an initial envelope mass of 0, 0.01, and 0.1 M$_{\oplus}$, respectively. Curves with high and low opacity represent the cases when the oxygen does or does not react with the surface, respectively. {\it Second row:} O$_{2}$ sink in the mantle. {\it Third row:} O$_{2}$ accumulated in the atmosphere {\it Fourth row:} Crossover mass. The horizontal light blue line represents the limit where the oxygen can be dragged along the hydrogen background out of the atmosphere.}
    \label{fig:WaterLossEvol}
\end{figure*}

\autoref{fig:WaterLossEvol} presents the atmospheric escape results for an initial water mass of 1, 3, and 10 terrestrial oceans (TO) 
under various initial hydrogen envelope masses. The results show that both planets can lose up to 9.9 TO in the first 370 Myr if they did not possess a primordial hydrogen envelope. In these scenarios, when the initial amount of water is 10 TO, the oxygen in the atmosphere can reach 1900 bars. In the early stage of evolution, during the first 1 Gyr, a small hydrogen envelope of 0.01 - 0.1 M$_{\oplus}$ is sufficient to prevent water escape.

Another feature that helps keep the water on the surface is the presence of oxygen in the atmosphere. Since oxygen can be dragged into space by the hydrogen, less energy is available for photolysis and the kinetic energy of the hydrogen atoms. If we compare all the panels in the first row from Figure \ref{fig:WaterLossEvol}, we can notice that for the same age, the planets lose more water if the oxygen is absorbed by the surface, compared to when the oxygen remains in the atmosphere.

\begin{figure*}
    \centering
    \includegraphics[width=0.9\linewidth]{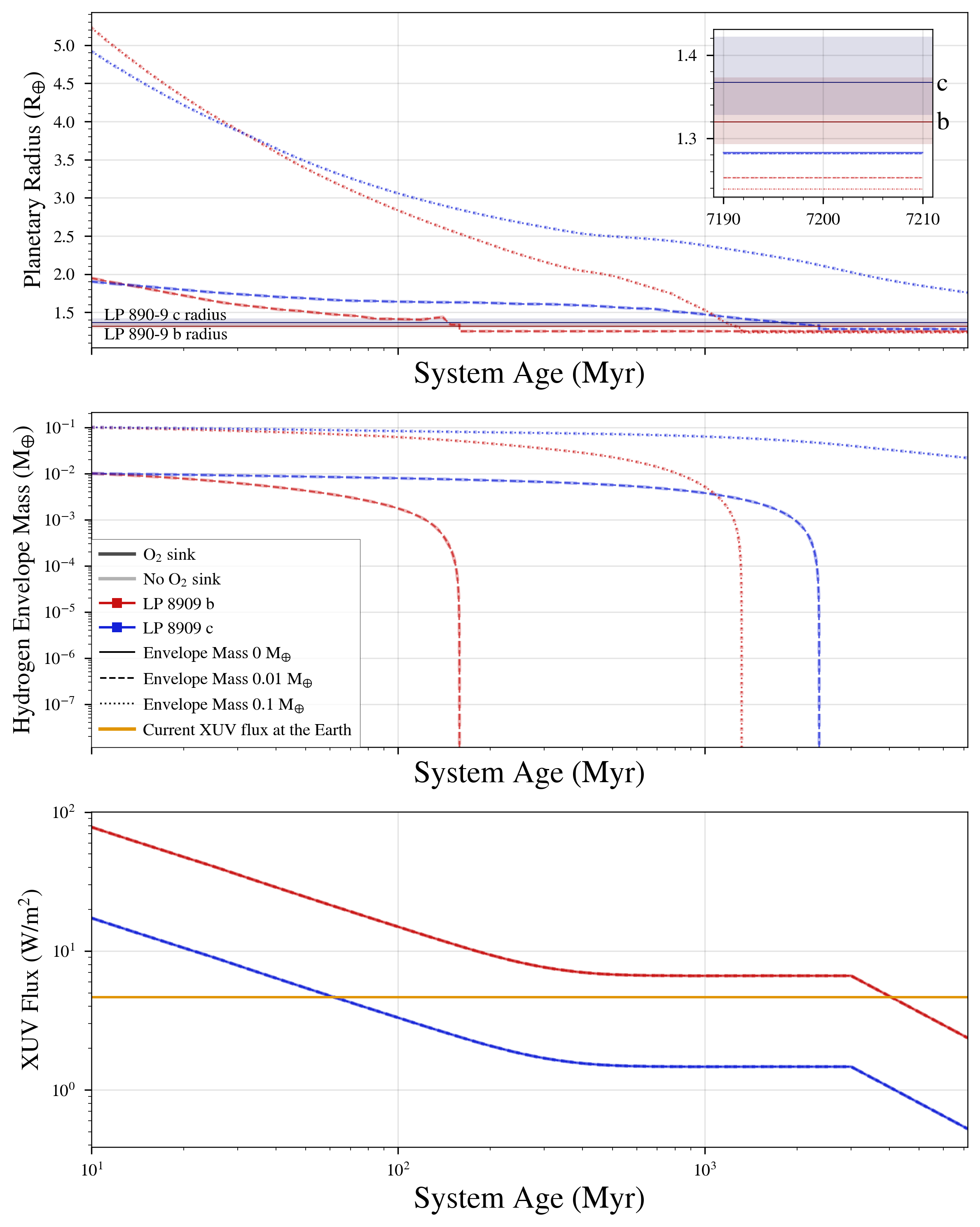}
    \caption{Time evolution of planetary radius (upper panel), hydrogen envelope mass (middle panel), and XUV flux at the planetary atmosphere (bottom panel). The horizontal orange line in the bottom panel indicates the XUV flux on the present Earth. Horizontal lines in the top panel indicate the observed planetary radii of LP 890-9 b and c \citep{Delrez22}. The shadow in the inset of the top  panel shows the planetary radius uncertainty.}
    \label{fig:WaterLossParams}
\end{figure*}

These figures reveal that envelope masses of $0.1\mearth$ and larger cannot be fully stripped during the pre-main sequence phase of the host star. These envelopes inflate the radius of the planet beyond the observed value and are therefore disfavored. However, after 7.5 Gyr of evolution, the results shown in the middle panel of \autoref{fig:WaterLossParams} indicate that these planets could have started with an H/He inventory in between 0.01-0.1 M$_{\oplus}$.

The results also show that the incoming XUV flux is initially hundreds of W/m$^{2}$, but drops quickly (bottom panel of \autoref{fig:WaterLossParams}). Due to this high XUV flux, the crossover mass at the beginning of the evolution is larger than 16 amu (fourth panel of \autoref{fig:WaterLossEvol}), which means that the hydrogen can drag the oxygen to space. But as the XUV flux drops, the crossover mass drops below 16 amu and there is no longer a flux of O atoms to space. The panels of the second row show the amount of oxygen that is absorbed by the surface. Planet b loses up to 67 bars and planet c loses up to 41 bars of oxygen to space when the planets begin with an initial surface water content of 10 TO. However, not all scenarios present extreme scenarios, such as total desiccation. Simulations with an initial hydrogen envelope mass of $>0.01$M$_{\oplus}$ can keep all the surface water, regardless of the initial amount of water on the surface.

\subsection{Orbital Evolution}

Next we turn to the orbital evolution of the system. We begin by examining the likelihood that the planets are in a 3:1 mean motion resonance. Of our 13,824 simulations, we found only 6 cases for which the resonance arguments librated, see Eqs.~(\ref{eq:resarg1})--(\ref{eq:resarg3}). Even then, the amplitudes were nearly 360$^\circ$, suggesting that the resonant effects are still very weak. We therefore conclude that we can safely ignore the 3:1 MMR and simulate the system with a secular orbital evolution model. 

\begin{figure*}
    \centering
    \includegraphics[width=\linewidth]{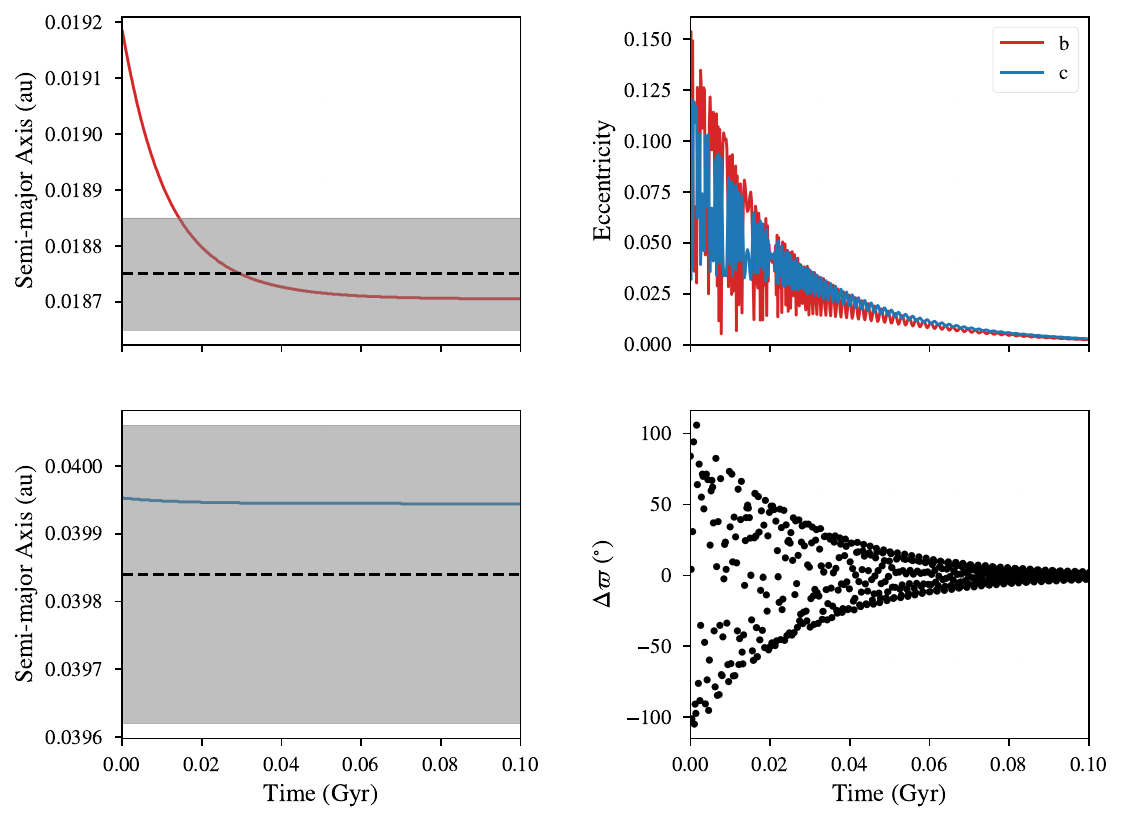}
    \caption{Example of the coupled tidal + orbital evolution of the LP 890-9 planetary system over 100 Myr. Dashed lines represent the best fit values and the shaded grey regions demarcate the $1\sigma$ credible intervals from \citet{Delrez22}.}
    \label{fig:orbit-evol}
\end{figure*}

Figure \ref{fig:orbit-evol} shows an example of the tidal-secular orbital evolution of the system, specifically the semi-major axes $a$,  eccentricities $e$, and the difference between their longitudes of periastron $\Delta\varpi$. This simulation reproduces the observed semi-major axes  within observational uncertainties \citep{Delrez22} after 100 Myr, but is very likely not the true trajectory because the magnitude and frequency dependence of the tidal dissipation model are poorly approximated by equilibrium tide theory \citep[see e.g.,][]{Touma1993,EfroimskyMakarov13,Greenberg09}. 

Tidal effects drive both eccentricities to $<0.01$ after $\sim 100$ Myr, while the perturbations between the planets drive sinusoidal oscillations. The result is effectively a damped-driven harmonic oscillator that ultimately circularizes both planets' orbits. Once the orbits have circularized, the tidal evolution of the semi-major axes can only proceed via dissipation in the star \citep{MardlingLin02,Raymond08}.

\subsection{Magma Ocean Evolution}

\begin{figure*}
    \centering
\includegraphics[width=\linewidth]{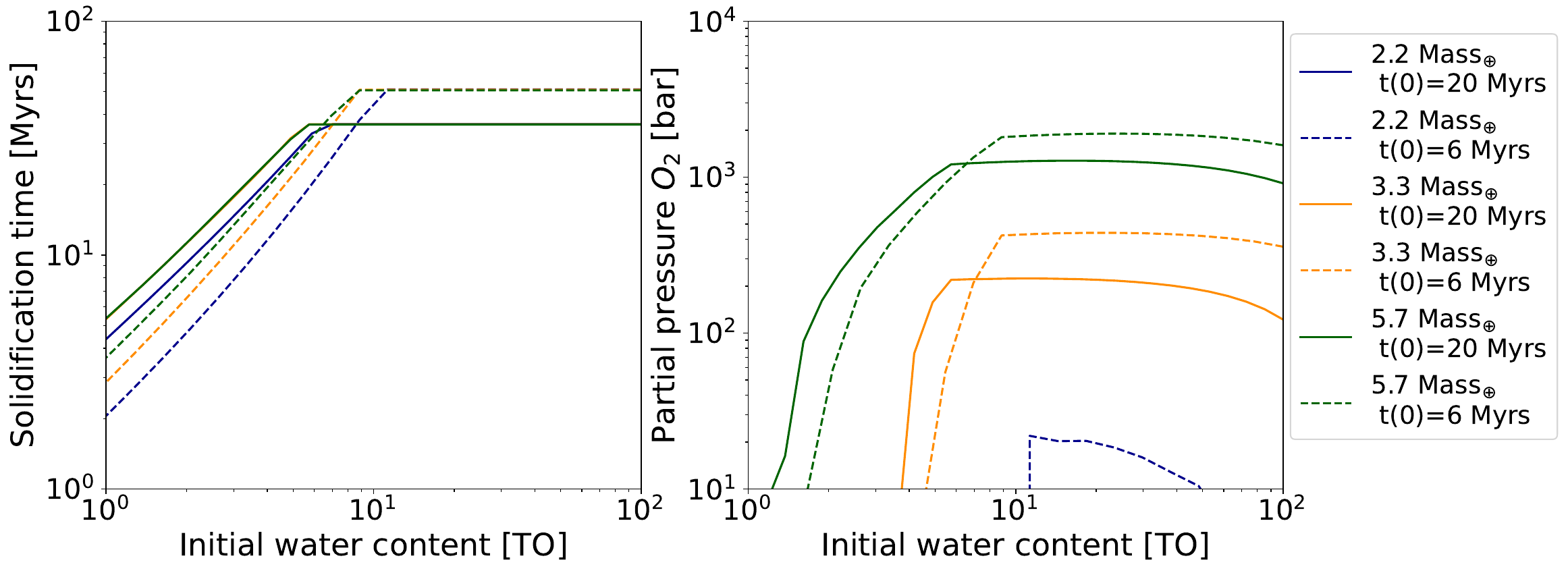}   
    \caption{{\it Left:} Magma ocean solidification time of LP 890-9 c versus initial water content in TO for different planetary masses and for disk dissipation times $t(0)$ of 20~Myrs (solid lines) and 6~Myrs (dashed lines), respectively. (We note that the solid orange line lies underneath the solid green line.) {\it Right:} Abiotic oxygen build-up (partial pressures of $O_2$ in bar) in LP 890-9 c during the magma ocean stage as a function of initial water mass.}
    \label{fig:magmoc_solidTime}
\end{figure*}

We find that the duration of the magma ocean phase depends on the initial water content and also on the assumed disk dissipation time that determines the start of our magma ocean evolution, $t(0)$. Assuming a disk dissipation time of 6~Myrs results in solidification times of 51~Myrs for large initial water content ($>9~TO$). If the disk dissipates later, that is at 20~Myrs, the maximum solidification is lowered to 36~Myrs. This tendency is a consequence of the changes in cooling rate that is determined by the difference between the outgoing long wave radiation (OLR) of a steam-dominated atmosphere and the irradiance of the host star. The OLR is always 282~W/m$^{2}$~\citep[e.g.][]{Barth21,Goldblatt15}, but the irradiance from the host star is lower at an age of 20~Myrs compared to 6~Myrs, resulting in faster magma ocean solidification for $t(0)=20$~Myrs (Fig.~\ref{fig:magmoc_solidTime}, left).

For low water content ($<6$~TO), the opposite is true. The solidification time is shorter when assuming fast disk dissipation (6~Myrs) compared to slow dissipation scenarios (20~Myrs). Here the magma ocean solidifies as the atmosphere desiccates via XUV photolysis of H$_2$O. Since the host star emits more XUV at 6~Myrs compared to 20~Myrs, desiccation, and thus solidification, occurs faster. The more efficient erosion for a younger host star is also evidenced by complete desiccation occuring for only up to 6~TO initial water with $t(0)=20$~Myrs and up to 9~TO with $t(0)=6$~Myrs. Generally, we find that the simulated LP~890-9 c magma ocean evolution resembles that of TRAPPIST-1~e as outlined in \citet{Barth21}, which is also close to the inner edge of the habitable zone of its host star.

\begin{figure*}
    \centering
    \includegraphics[width=\linewidth]{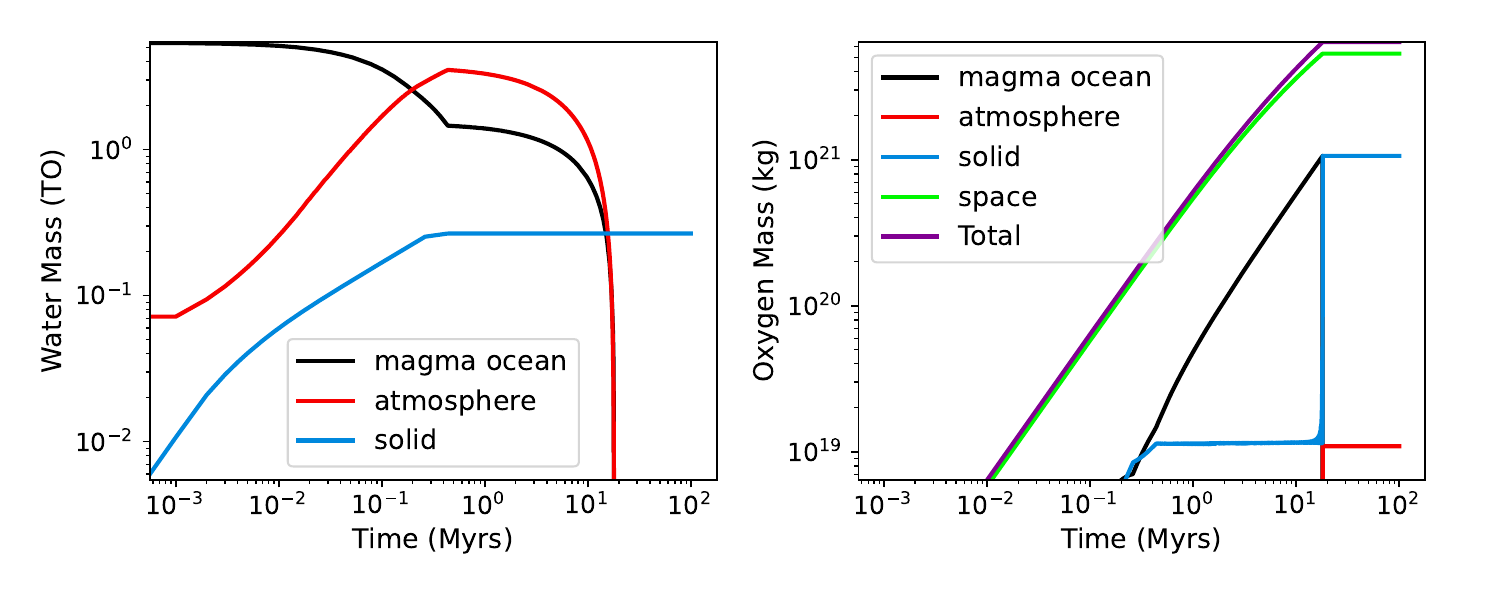}
    \caption{Example simulation of LP 890-9 c assuming a mass of $2.2\, \mathrm{M_\oplus}$ and 5 TO of initial water content, assuming disk dissipation at 6~Myrs. {\it Left:} Water evolution during the magma ocean stage shows all water is lost from the atmosphere after 18 Myr. {\it Right:} Oxygen evolution during the magma ocean stage, where the majority of the O$_2$ escapes into space. }
    \label{fig:magmoc_5TO}
\end{figure*}

Water loss leads to the abiotic build-up of oxygen in the atmosphere once the iron buffer in the mantle has been fully exhausted or when the mantle fully solidifies (Fig.~\ref{fig:magmoc_solidTime}, right). We find that the oxygen surface pressures can range from 400~bars ($3.3 M_{Earth}$) to 2000~bars ($5.7 M_{Earth}$) for disk dissipation times of 6~Myrs. Longer dissipation times result in a moderate reduction of O$_2$ build-up by less than an order of magnitude due to reduced overall water erosion at this later stellar evolution stage. For the lowest planetary mass, $2.2 M_{Earth}$, oxygen build-up is suppressed for all disk dissipation scenarios as we find that the majority of the oxygen produced by photolysis of H$_2$O is lost to space. Figure~\ref{fig:magmoc_5TO} shows the abundances of water and oxygen in the various reservoirs over time for an initial water mass of 5 TO. 

\subsection{Thermal and Magnetic Evolution}

\begin{figure*}
    \centering
    \includegraphics[width=\linewidth]{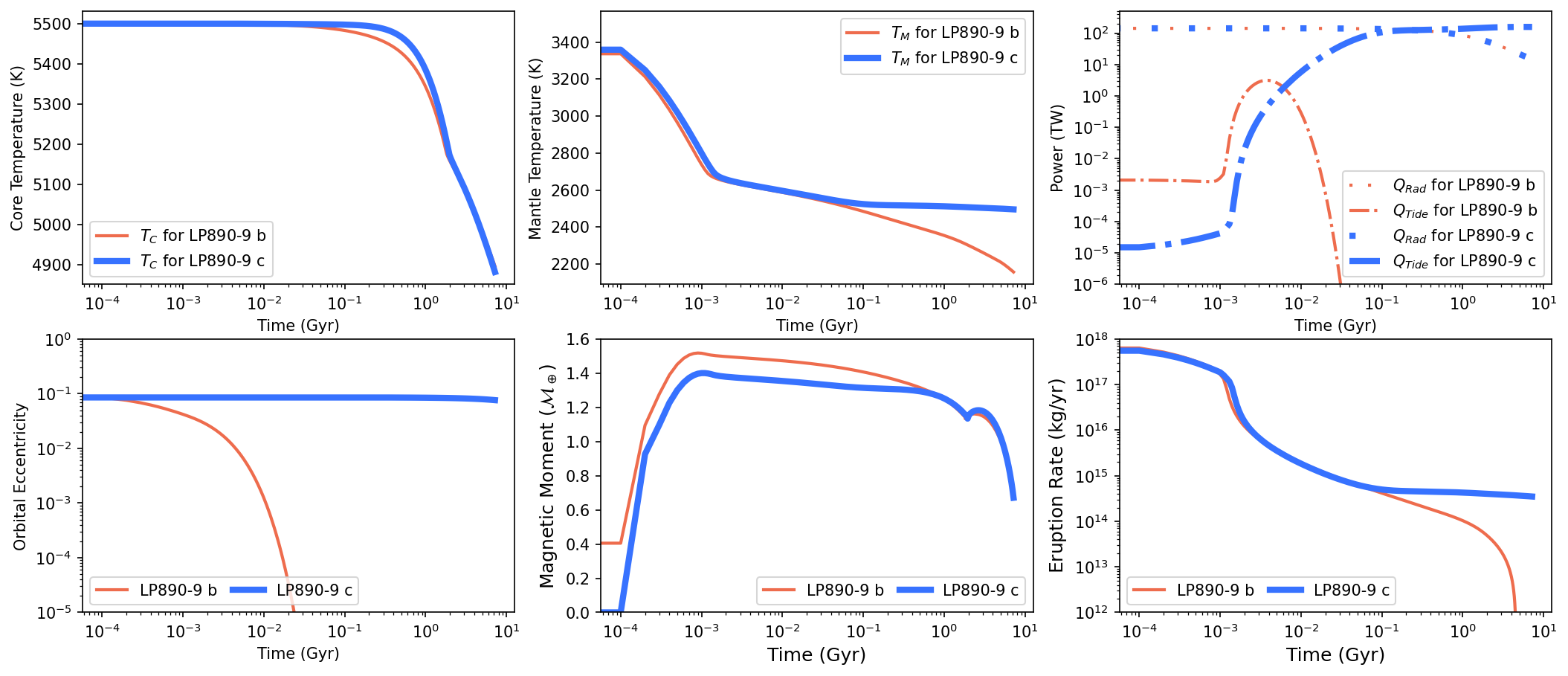}
    \caption{An example of the thermal and magnetic evolution of LP 890-9 b and c from the \eqtide, \thermint, and \radheat modules. Both fiducial integrations are representative cases from the larger integration set presented in Figure \ref{fig:thermmag_energysource}, with the viscosity activation energies of $A_{v, b} = 2.9\cdot 10^{5}$ and $A_{v, c} = 3.1\cdot 10^{5}$ chosen so that both planets have approximately equal initial tidal $Q \approx 30$. Both planets have initial eccentricities of $e\sim0.1$, and masses of $m_{b} = 2.3 M_{\oplus}$ and $m_{c} = 2.9 M_{\oplus}$. The interior structures of both planets are otherwise assumed to be Earth-like. Throughout the integration, LP 890-9 c has a consistently higher volumed-averaged core temperature and generally has a higher  volumed-averaged mantle temperature than LP 890-9 b. In the upper right panel, we see that LP 890-9 b's internal heat budget is driven by radiogenic heating for the full integration lifetime, while LP 890-9 c has a significant contribution from tidal heating, which increases in total power as its orbital eccentricity begins to decay by the end of the integration (bottom left panel). The tidal heating drives mantle heating, leading to higher sustained volume-averaged mantle temperatures for LP-890 c. Both planets have non-zero magnetic moments for the majority of the integration, and experience declining eruption rates as the planets age.}
    \label{fig:thermmag_fiducialrun}
\end{figure*}

A fiducial case of the thermal and magnetic evolution of planets b and c is shown in Figure \ref{fig:thermmag_fiducialrun}. While the exact parameters at 7.2 Gyr will change depending on the initial parameters in mass, eccentricity, and mantle viscosity activation energy, the general evolution will be similar to that shown in Figure \ref{fig:thermmag_fiducialrun}. From this fiducial run, we can see that, despite having similar planetary masses, the differing levels of tidal heating between the two planets result in significant differences in their thermal evolution. 

For LP 890-9 b, the orbit circularizes quickly (bottom left panel), resulting in almost no contribution to the internal heat budget from tidal heating (top right panel). In contrast, LP 890-9 c experiences tidal heating that reaches 1 TW at 10 Myr, and $\sim$100 TW at 10 Gyr (Earth's internal power is ~40 TW). The impact of this powerful heating drives sustained higher mantle temperatures of $\sim2500$ K for planet c. Planet b, in contrast, continues to cool to $T_m \sim 2200$ K by the end of the integration. In the top middle panel, we see the mantle temperatures begin to diverge significantly at $\sim 0.1$ Gyr, reaching a maximum difference between the two planets of about 300 K at $\sim5$ Gyr.

This relatively small difference in volumed-averaged mantle temperature creates a qualitative difference in the core temperature evolutions. In the top left panel we see that up to 0.05 Gyr, the volumed-averaged core temperatures are nearly equal and constant, but that afterwards b's core temperature decreases while c's increases by about 30 K before both rapidly falling off at about 1-2 Gyr. This divergence is due to the competing effects of a hotter mantle and radiogenic heating from potassium (half-life = 1.8 Gyr) in the core. Initially the cores are in thermodynamic equilibrium with the mantles as core cooling into the mantle is approximately offset by radiogenic heating. As planet c's mantle power increases due to tidal heating, the temperature difference decreases, resulting in slower core cooling. Thus, the core heats up by about 30 K. After 1-2 Gyr, potassium heating fades and the mantles continue to cool, both of which increase core cooling and instigate the steeper temperature drop in the core seen in the top left panel of Fig.~\ref{fig:thermmag_fiducialrun}.

The different core and mantle thermal evolutions is also reflected in the magnetic field strengths. With a shallower temperature gradient between planet c's core and mantle, we would expect its liquid core to experience less vigorous convection and hence generate a weaker dynamo, which is confirmed in the bottom middle panel. In \texttt{VPLanet}'s 1D model, the dynamo is independent of $\dot{T_c}$, so the brief period in which the core warms does not affect dynamo generation. The discontinuity both magnetic moments show at later times is due to inner core nucleation, which provides additional core heating in the from of latent heat.

Although this thermal evolutionary model does not strictly include a magma ocean phase, the effects of a high mantle temperature are visible in Figure \ref{fig:thermmag_fiducialrun}. As the temperature drops, the melt fraction also drops and after 1 Myr the mantle viscosity increases dramatically as a result. Convection in the stiffer mantle slows cooling, and as a result the mantle temperature becomes approximately constant. We emphasize, though, that these simulations did not include the full geochemical evolution of the magma ocean, and so evolution during the first million years is only shown for context.

The top right panel of Figure \ref{fig:thermmag_fiducialrun} shows the relative importance of two major internal heat sources: radiogenic heating and tidal heating. Radiogenic heating is supplied by radioactive isotopes present in the planetary interior, and the rate of heating depends only on the abundance of those isotopes. In contrast, tidal heating occurs due to tidal strain dissipated in the planetary interior, and is a function of $e$ and rotation period. For the fiducial case given in Figure \ref{fig:thermmag_fiducialrun}, the orbit of the inner planet LP 890-9 b circularizes well before the present day, resulting in the internal heat budget being fully provided by radiogenic heating. In contrast, the outer planet LP 890-9 c still maintains significant planetary eccentricity by a stellar age of 7.2 Gyr, meaning that the internal heat budget comes from multiple sources. 

\begin{figure}
    \centering
    \includegraphics[width=\linewidth]{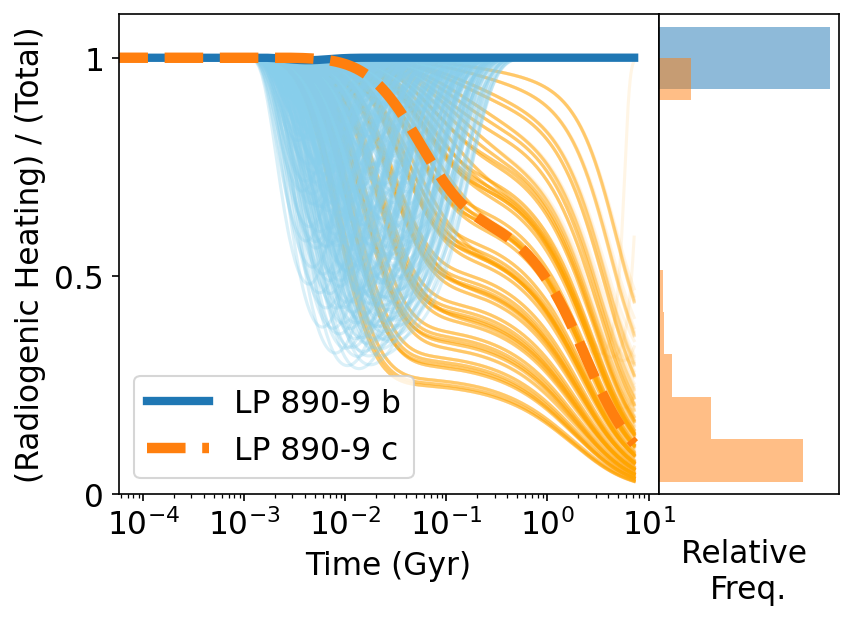}
    \caption{The results of a parameter sweep over the mass, orbital eccentricity, and viscosity activation energy for each planet. Individual trials are plotted as thin solid curves. The evolution of the fiducial case is overlaid in the left panel with thick curves. The distribution of the fraction of the total internal heat generated by radiogenic heating at the present day (a stellar age of 7.2 Gyr) is plotted as a histogram for each planet in the right panel. The orbit of the inner planet, LP 890-9 b, always circularizes by the end of the integration, leaving the primary source of internal heating to be radiogenic heating in all integrations. The outer planet, LP 890-9 c, has a much more varied range of heating ratios at the end of the integration, so that the primary internal heat source at present day cannot be securely determined.}
    \label{fig:thermmag_energysource}
\end{figure}

We ran integrations of the system's interior evolution over these bounds, and in Figure \ref{fig:thermmag_energysource} we plot the ratio of the radiogenic heating to the total heating (when the total includes contributions from radiogenic and tidal heating). The two planets have different bounds of potential evolution: LP 890-9 b shows evidence of some contribution from tidal heating at early times, but by 7.2 Gyr all system realizations approach a ratio of 1; in contrast, LP 890-9 c allows a larger range of values of the ratio, with the most likely value at 7.2 Gyr being 0. From these parameter sweeps, we can conclude that LP 890-9 b's internal heat budget is supplied solely by radiogenic heating today. In contrast, while LP 890-9 c's heat budget can be supplied by solely radiogenic heating, predominantly tidal heating, or by a combination of the two. 

\section{Discussion}

The results presented here reveal numerous possibilities for the evolution of the LP 890-9 system and the potential habitability of planet c. Crucially, we find no reason to conclude at this time that LP 890-9 c cannot support life. Of course, there are many reasons to still be cautious about that possibility, and the research presented herein illuminates many of those reasons, as well as valuable directions of future research that could resolve the ambiguities.

Starting with the star, the lack of knowledge of the current XUV luminosity is a major missing piece for assessing habitability. As the incident XUV flux is a primary driver of water destruction \citep{Watson81,Erkaev07,LugerBarnes15}, observations of this parameter are probably the most important for assessing habitability. The XUV evolution could potentially be further constrained if the stellar rotation rate were measured, which can provide constraints on the stellar convective Rossby number, which is tightly correlated with activity and high energy radiation \citep[\eg][]{Barnes03,Garraffo18,Johnstone21}. Combining a model of stellar rotation evolution with the observed XUV luminosity could provide much stronger constraints on the XUV saturation time and fraction, as well as the subsequent XUV luminosity decay.

Moving onto the planet itself, the biggest missing piece is the planetary mass. Without the mass, we cannot estimate the gravitational acceleration at the surface, which is a key controller of atmospheric escape. If future radial velocity or transit timing variations \citep{Agol05,HolmanMurray05} could weigh the planets and add that constraint to the models presented here, then they would be significantly more accurate.

Measuring the planetary mass could also provide clues to planetary composition. Here we have assumed an Earth-like composition, which influences processes such as internal temperature, viscosity, magnetic field strength, and tidal effects. While a comprehensive understanding of how composition affects the full range of planetary processes is not yet available, knowledge of the composition could nonetheless help constrain the thermal and volatile evolution of the planets.

We find that planet c could lose 10 TO of water due to thermal escape, consistent with other results for planets orbiting low mass stars \citep{LugerBarnes15,Bolmont17,Amaral22}. The water loss process could result in hundreds of bars of oxygen in the atmosphere or chemically absorbed by the surface. Alternatively, just 0.001$M_\oplus$ of primordial hydrogen could protect the planet's water and allow the planet to emerge as a ``habitable evaporated core'' after the hydrogen envelope is completely lost \citep{Luger15,Amaral22}. We caution, however, that our models are relatively simple and future work could use more sophisticated models to further refine water loss predictions \citep{Cohen15,Brain16,Dong18,Airapetian20}.

We find the orbital evolution is probably not a major influence on the potential habitability of planet c. Although we could not determine if the planets are in a 3:1 MMR, its effects are likely to be weak because the period ratio is significantly different from the exact commensurability. The long-term tidal/orbital evolution is not likely to result in significant semi-major axis evolution, at least for initial eccentricities below 0.2. Previous work has shown that the planetary spins are likely to synchronize within a few tens of kyr \citep{Barnes17}, and  the obliquities damp to equilibrium values that are well below 1$^\circ$ \citep{Dobrovolskis09,Heller11}, and hence the planets are likely only illuminated on one hemisphere. 

\section{Conclusions}

The LP 890-9 system offers an enticing opportunity to study planetary evolution and habitability around low mass stars. The presence of an Earth-sized planet in the HZ is sufficient to warrant further scrutiny, and our analysis has provided more reasons to turn large space-borne telescopes toward this system. Regardless of habitability, the planetary atmospheres could provide clues to the atmospheric, internal, and orbital evolution of any terrestrial planet orbiting a low-mass star.

The most serious threats to habitability are a persistent or transient runaway greenhouse, which ultimately results in the destruction of water. The early pre-main sequence brightness of the star could desiccate planet c before it ever reaches the habitable zone \citep{LugerBarnes15}, or a more recent ``tidal greenhouse'' \citep{Barnes13,DriscollBarnes15} could permanently destroy habitability at later times. While these threats are real, our analysis cannot confirm that they occurred, hence the ability of LP 890-9 c to support life remains an open question.
However, the LP 890-9 system adds to the limited sample of objects available for investigating demographic imprints of the inner boundary of the HZ~\citep{Turbet2019,Schlecker2024}.

Our investigation has confirmed that future research should focus on measuring the current XUV luminosity of the star and the planetary masses. X-ray and UV measurements could be made that are similar to TRAPPIST-1 \citep{Becker20}, while transit timing variations could reveal their masses \citep[see, \eg][]{Agol21}. Furthermore, if the composition of the star could be measured and/or the processes that formed these worlds be determined, then the composition of the planets could be constrained \citep{Bond10}.

Future research should also improve upon the modeling efforts described above. While our study has examined many aspects of planetary evolution, it is still deficient in many areas. In particular, the role of carbon dioxide in the interior and atmosphere must be addressed \citep[see \eg][]{KrissansenTottonFortney22}. As CO$_2$ is a heavy molecule and a powerful greenhouse gas, it could accumulate in the atmosphere for Gyr and help heat the surface into a runaway greenhouse. Ultimately, a model that connects the magma ocean, stagnant lid/plate tectonics geochemistry to orbital evolution and improved XUV histories will provide better insight into planet c's habitability prior to direct atmospheric characterization. \vplanet possesses many of the pieces of such a model, but significant effort will be required to properly couple and validate the physics and chemistry.

As one of just a few potentially habitable planets that transits stars in the solar neighborhood, LP 890-9 c is a very valuable target in the search for life beyond the Solar System. While our research cannot provide many definitive statements about the history and habitability of the LP 890-9 system, it has revealed numerous intriguing possibilities and clarified directions of future research. Our investigations provide a starting point for further characterizing this system, and perhaps the first discovery of active biology on an exoplanet.

\vspace{0.5cm}

This work represents the output of the Second VPLanet Workshop held 12-13 Sep 2022. RB, RG, and MG acknowledge support from NASA grant number 80NSSC20K0229 and the NASA Virtual Planetary Laboratory Team through grant number 80NSSC18K0829. JB, RG, and MG acknowledge support from NSF graduate student fellowships.  LNRA acknowledges the support of UNAM DGAPA PAPIIT project IN110420 and thanks CONACYT’s graduate scholarship program for its support. JCB has been supported by the Heising-Simons \textit{51 Pegasi b} postdoctoral fellowship.  BALG acknowledges support from a Chalmers Cosmic Origins postdoctoral fellowship.  MAS acknowledges the support from the Student Research Award in Planetary Habitability by the Center for Planetary Systems Habitability at the University of Texas at Austin. LC acknowledges funding by the Royal Society University Fellowship URF~R1~211718. The results reported herein benefited from collaborations and/or information exchange within NASA’s Nexus for Exoplanet System Science (NExSS) research coordination network sponsored by NASA’s Science Mission Directorate under Agreement No. 80NSSC21K0593 for the program ``Alien Earths''. Finally, we thank two anonymous referees who greatly improved the quality and clarity of this manuscript.

\bibliography{lp890}

\end{document}